# SPECTRAL ANALYSIS OF DISCRETE-TIME QUANTUM WALKS IN THE QUARTER PLANE

**CLEMENT AMPADU**

31 Carrolton Road
Boston, Massachusetts, 02132
USA
e-mail: drampadu@hotmail.com

**Abstract**

Using the Cantero-Grunbaum-Moral-Velazquez (CGMV) method, we obtain the spectral measure for the quantum walk.

I. Introduction

We study the discrete-time quantum walk in the half-plane by means of spectral analysis. The analysis is based on the CGMV method introduced by Cantero et.al [1]. The CGMV method used here is based on the spectral theory on the unit circle, which has been studied by a number of authors, for examples, see [20-24]. Apart from the CGMV method, analysis of discrete-time quantum walks by path counting method and the Fourier analysis, collectively, is popular [2-19].

Let $\mu$ be a probability measure on $\partial D = \{z \in C : |z| = 1\}$, where $C$ is the set of complex numbers, and let $L^2_\mu(D)$ denote the Hilbert space of $\mu$-square integrable functions on $\partial D$. The Laurent polynomials $\{x_l(z)\}_{l=0}^{\infty}$ are orthogonal polynomials on $\partial D$ obtained by applying the Grahm-Schmidt orthonormalization to $\{1, z, z^{-1}, z^2, z^{-2}, \cdots\}$ with respect to the inner product

$(f, g) = \int_{z \in \partial D} \overline{f(z)} g(z) d\mu(z)$, where $f(z), g(z) \in L^2_\mu(\partial D)$. A recurrence relation for the Laurent polynomials can be obtained via the recurrence relation for Szego polynomials and the relation between the Laurent and Szego polynomials themselves [22,23]. The simplest expression for the recurrence relation of the Laurent polynomials is given by the CMV matrix, which is a five-diagonal band representation associated with the Verblunsky parameter $(\alpha_0, \alpha_1, \alpha_2, \ldots)$ [24]. In this paper we consider the following two cases: (i) the CVM matrix with null-odd Verblunsky parameter, that is, $(\alpha_0, \alpha_1, \alpha_2, \ldots) = (\alpha, 0, \alpha, 0, \ldots)$ and (ii) the CVM matrix with null-even parameter, that is, $(\alpha_0, \alpha_1, \alpha_2, \ldots) = (0, \alpha, 0, \alpha, \ldots)$.

We show, in this paper using the CGMV method that localization of the quantum walk derives from the point mass of the spectral measure on $\partial D$. We also give explicit expressions for the limit measures, and the necessary and sufficient condition for localization.

The rest of this paper is organized as follows. In Section II, we explain the relation between the CVM matrix and the corresponding quantum walk. In Section III, we consider the relation between the spectral measures of the CVM matrix and localization of the corresponding quantum walk. Section IV presents the main results with proof. The limit measure of Type I and Type II quantum walks is given in Theorems 1 and 3, respectively. The necessary and sufficient condition for localization is given in Corollary 2 and Corollary 4 for Type I and Type II quantum walks, respectively. The first condition for the necessary and sufficient condition for localization in Type I quantum walks appearing in Corollary 2 depends only on the quantum coin whilst the second condition appearing in the same Corollary depends not only on the quantum coin but also the initial coin state. For Type II quantum walks, the necessary and sufficient condition for localization given by Corollary 4, is independent of the initial state $\begin{bmatrix} e^{i\delta_1} & e^{i\delta_2} \end{bmatrix}^T$. Section V is devoted to the summary and open problem.

## II. Quantum walks governed by Cantero-Moral-Velazquez (CMV) matrices

Let $C_{(\alpha_0,\alpha_1,\cdots)}$ be the CMV matrix associated with the Verblunsky parameter $(\alpha_0,\alpha_1,\cdots)$. The simplest matrix representation for $C_{(\alpha_0,\alpha_1,\cdots)}$ is given by the following form [20,24]:

$$C_{(\alpha_0,\alpha_1,\cdots)} = \begin{bmatrix} \overline{\alpha}_0 & \rho_0\overline{\alpha}_1 & \rho_0\rho_1 & 0 & 0 & 0 & 0 & 0 & \cdots \\ \rho_0 & -\alpha_0\overline{\alpha}_1 & -\alpha_0\rho_1 & 0 & 0 & 0 & 0 & 0 & \cdots \\ 0 & \rho_1\overline{\alpha}_2 & -a_1\overline{\alpha}_2 & \rho_2\overline{\alpha}_3 & \rho_2\rho_3 & 0 & 0 & 0 & \cdots \\ 0 & \rho_1\rho_2 & -\alpha_1\rho_2 & -\alpha_2\overline{\alpha}_3 & -\alpha_2\rho_3 & 0 & 0 & 0 & \cdots \\ 0 & 0 & 0 & \rho_3\overline{\alpha}_4 & -\alpha_3\overline{\alpha}_4 & \rho_4\overline{\alpha}_5 & \rho_4\rho_5 & 0 & \cdots \\ 0 & 0 & 0 & \rho_3\rho_4 & -\alpha_3\rho_4 & -\alpha_4\overline{\alpha}_5 & -\alpha_4\rho_5 & 0 & \cdots \\ \vdots & \vdots & \vdots & \vdots & \vdots & & \ddots & & \end{bmatrix}$$

where $\alpha_j \in C$ satisfies $|\alpha_j| < 1$ and $\rho_j = \sqrt{1-|\alpha_j|^2}$. Let the *quantum coin* be given by

$$U = \begin{bmatrix} c_{RR} & c_{LR} & c_{UR} & c_{DR} \\ c_{RL} & c_{LL} & c_{UL} & c_{DL} \\ c_{RU} & c_{LU} & c_{UU} & c_{DU} \\ c_{RD} & c_{LD} & c_{UD} & c_{DD} \end{bmatrix}.$$ For $i,j \in \{L,R,D,U\}$, define $\rho = |c_{ii}|$, $\arg(c_{ii}) = \sigma_i$, $\det U = \Delta$,

where $\arg(z)$ is the argument of $z \in C$. By the unitary of $U$, $\Delta = e^{i\left(\sum_{j\in\{L,R,D,U\}}\sigma_j\right)}$, and for $i \neq j$, $c_{ij} = -\Delta \overline{c}_{ji}$. In this paper we will consider two type of quantum walks, Type I QW and Type II QW.

**Definition I (Type I QW):** The total space $H^{(I)}$ for this type of quantum walk is generated by a standard basis $\{|0,0,R\rangle, |0,0,L\rangle, |0,0,U\rangle, |0,0,D\rangle, |1,0,R\rangle, |0,1,U\rangle, \cdots\}$. The time evolution $W^{\{I,U\}}$ with quantum coin $U$ is denoted by

$$W^{(I,U)}|0,0,R\rangle = c_{RR}|1,0,R\rangle + c_{LR}|0,0,L\rangle + c_{UR}|0,1,U\rangle + c_{DR}|0,0,D\rangle$$

$$W^{(I,U)}|0,0,L\rangle = c_{RL}|1,0,R\rangle + c_{LL}|0,0,L\rangle + c_{UL}|0,1,U\rangle + c_{DL}|0,0,D\rangle$$

$$W^{(I,U)}|0,0,U\rangle = c_{RU}|1,0,R\rangle + c_{LU}|0,0,L\rangle + c_{UU}|0,1,U\rangle + c_{DU}|0,0,D\rangle$$

$$W^{(I,U)}|0,0,D\rangle = c_{RD}|1,0,R\rangle + c_{LD}|0,0,L\rangle + c_{UD}|0,1,U\rangle + c_{DD}|0,0,D\rangle$$

$$W^{(I,U)}|x,y,R\rangle = c_{RR}|x+1,y,R\rangle + c_{LR}|x-1,y,L\rangle + c_{UR}|x,y+1,U\rangle + c_{DR}|x,y-1,D\rangle \quad (x,y \geq 1)$$

$$W^{(I,U)}|x,y,L\rangle = c_{RL}|x+1,y,R\rangle + c_{LL}|x-1,y,L\rangle + c_{UL}|x,y+1,U\rangle + c_{DL}|x,y-1,D\rangle \quad (x,y \geq 1)$$

$$W^{(I,U)}|x,y,U\rangle = c_{RU}|x+1,y,R\rangle + c_{LU}|x-1,y,L\rangle + c_{UU}|x,y+1,U\rangle + c_{DU}|x,y-1,D\rangle \quad (x,y \geq 1)$$

$$W^{(I,U)}|x,y,D\rangle = c_{RD}|x+1,y,R\rangle + c_{LD}|x-1,y,L\rangle + c_{UD}|x,y+1,U\rangle + c_{DD}|x,y-1,D\rangle \quad (x,y \geq 1)$$

**Definition II (Type II QW):** The total space $H^{(II)}$ for this type of quantum walk is generated by a standard basis $\{|0,0,L\rangle, |0,0,D\rangle, |1,0,R\rangle, |0,0,D\rangle, |0,1,U\rangle, \cdots\}$. Let $\gamma_1, \gamma_2$ be real numbers. The time evolution $W^{\{II,U\}}$ with quantum coin $U$ is denoted by

$$W^{(II,U)}|0,0,L\rangle = e^{i\gamma_1}|1,0,R\rangle + e^{i\gamma_2}|0,1,U\rangle$$

$$W^{(II,U)}|0,0,D\rangle = e^{i\gamma_1}|1,0,R\rangle + e^{i\gamma_2}|0,1,U\rangle$$

$$W^{(II,U)}|x,y,R\rangle = c_{RR}|x+1,y,R\rangle + c_{LR}|x-1,y,L\rangle + c_{UR}|x,y+1,U\rangle + c_{DR}|x,y-1,D\rangle \quad (x,y \geq 1)$$

$$W^{(II,U)}|x,y,L\rangle = c_{RL}|x+1,y,R\rangle + c_{LL}|x-1,y,L\rangle + c_{UL}|x,y+1,U\rangle + c_{DL}|x,y-1,D\rangle \quad (x,y \geq 1)$$

$$W^{(II,U)}|x,y,U\rangle = c_{RU}|x+1,y,R\rangle + c_{LU}|x-1,y,L\rangle + c_{UU}|x,y+1,U\rangle + c_{DU}|x,y-1,D\rangle \quad (x,y \geq 1)$$

$$W^{(II,U)}|x,y,D\rangle = c_{RD}|x+1,y,R\rangle + c_{LD}|x-1,y,L\rangle + c_{UD}|x,y+1,U\rangle + c_{DD}|x,y-1,D\rangle \quad (x,y \geq 1)$$

In this paper we restrict initial states of Types I and II QW's to

$\Psi_0^I = \alpha|0,0,R\rangle + \beta|0,0,L\rangle + \mu|0,0,U\rangle + \zeta|0,0,D\rangle$ and $\Psi_0^{II} = e^{i\delta_1}|0,0,L\rangle + e^{i\delta_2}|0,0,D\rangle$,

respectively, where $|\alpha|^2 + |\beta|^2 + |\mu|^2 + |\zeta|^2 = 1$, and $\delta_1, \delta_2 \in \mathfrak{R}$

Define a total weight of the passage from position $(x_1, y_1)$ to position $(x_2, y_2)$ at time $t$ with

quantum coin $U$ by

$$\Xi^{(J,U)}_{(x_2,y_2),(x_1,y_1)}(t) = \sum_{d_1,d_2,d_3,d_d \in \{R,L,U,D\}} I_{\{((x_2,y_2),(d_1,d_2)),((x_1,y_1),(d_3,d_4))\in H^J\}}((d_1,d_2),(d_3,d_4))$$
$$\times \langle x_2,y_2,d_1,d_2|(W^{(J,U)})^t|x_1,y_1,d_3,d_4\rangle |d_1,d_2\rangle\langle d_3,d_4| \qquad (J \in \{I,II\})$$

where $|R\rangle = [1,0,0,0]^T, |L\rangle = [0,1,0,0]^T, |U\rangle = [0,0,1,0]^T, |D\rangle = [0,0,0,1]^T$, and

$$I_A(x,y) = \begin{cases} 1, \text{if } (x,y) \in A \\ 0, \text{otherwise} \end{cases}$$

Let $(X_t^{(I)}, Y_t^{(I)})$ and $(X_t^{(II)}, Y_t^{(II)})$ be Type I and II quantum walks respectively, then

$$P((X_t^{(J,U)}, Y_t^{(J,U)}) = (x,y)) = \left\| \Xi^{(J)}_{x,y,0,0} \varphi_0^{(J)} \right\|^2, \text{ where } \varphi_0^{(J)} = \begin{cases} [\alpha \ \beta \ \mu \ \zeta]^T, \text{if } J = I \\ [e^{i\delta_1} \ e^{i\delta_2}]^T, \text{if } J = II \end{cases}. \text{ To get}$$

the matrix representation of $W^{(I,U)}$ we give a 1-1 correspondence between the basis of $H^{(I)}$ and $\{0,1,2,3,\cdots\}$ such that

$(0,0,R) \leftrightarrow 0; (0,0,L) \leftrightarrow 1; (0,0,U) \leftrightarrow 2; (0,0,D) \leftrightarrow 3; (k,k,R(U)) \leftrightarrow 2k;$

$(k,k,L(D)) \leftrightarrow 2k+1; (k \geq 1)$. Let $\Lambda_I$ be the diagonal matrix, where

$$(\Lambda_I)_{ij} = \delta_{i,j} \lambda_i^{(I)} = \begin{cases} 1, i = j = 0 \\ \lambda_i^{(I)}, \text{if } i = j \neq 0 \\ 0, \text{if } i \neq j \neq 0 \end{cases}, \text{ with } \lambda_{2k}^I = e^{-ik(\sigma_R + \sigma_U)} (k \geq 0), \lambda_{2k-1}^I = e^{-ik(\sigma_L + \sigma_D)} (k \geq 1).$$

Then we see that $W^{(I,U)} = \Lambda_I^T C_{(a_0,0,a_2,0,\cdots)} \Lambda_I^*$, where $a_j = \begin{cases} a\Delta^{\frac{-(j+1)}{2}}, j \text{ even} \\ 0, j \text{ odd} \end{cases}$, and

$$a = \left( \sum_{\substack{m,n \in \{L,R,D,U\} \\ m \neq n}} \overline{c}_{mn} \right) \Delta^{\frac{1}{2}}.$$

We should remark that this gives a relation between Type I QW and the null-odd CMV matrix, that is, the CMV matrix with null-odd parameter $(\alpha_0, \alpha_1, \alpha_2, \cdots) = (\alpha, 0, \alpha, 0, \cdots)$. For the Type II quantum walk, the one-to-one correspondence between the basis $H^{(II)}$ and $\{0,1,2,3,\cdots\}$ is such that $(0,0,L) \leftrightarrow 0$, $(0,0,D) \leftrightarrow 1$, $(k,k,R) \leftrightarrow 2k-1$, $(k,k,U) \leftrightarrow 2k-1$ $(k \geq 1)$. Define

$$(\Lambda_{II})_{ij} = \delta_{i,j} \lambda_i^{(II)} = \begin{cases} 1, i = j = 0 \\ \lambda_i^{(II)}, \text{if } i = j \neq 0 \\ 0, \text{if } i \neq j \neq 0 \end{cases}, \text{ with } \lambda_{2k}^{(II)} = e^{-ik((\sigma_L + \sigma_D) - (\gamma_1 + \gamma_2))}, \lambda_{2k+1}^{(II)} = e^{ik((\sigma_R + \sigma_U) - (\gamma_1 + \gamma_2))}.$$

Thus we that $W^{(II,U)} = e^{i(\gamma_1 + \gamma_2)} \Lambda_{II} C_{(0, b_1, 0, b_3, \cdots)} \Lambda_{II}^*$, where $b_j = \begin{cases} b(e^{-i(\gamma_1 + \gamma_2)}\Delta), j \text{ odd} \\ 0, j \text{ even} \end{cases}$ and $b = \left( \sum_{\substack{m,n \in \{L,R,D,U\} \\ m \neq n}} \overline{c}_{mn} \right) \Delta e^{-i(\gamma_1 + \gamma_2)}$. We should remark that $W^{(II,U)} = e^{i(\gamma_1 + \gamma_2)} \Lambda_{II} C_{(0, b_1, 0, b_3, \cdots)} \Lambda_{II}^*$ gives a relation between the null-even CMV matrix and Type II QW, that is, the CVM matrix with null-even parameter $(\alpha_0, \alpha_1, \alpha_2, \cdots) = (0, \alpha, 0, \alpha, \cdots)$. Denote the Laurent polynomials for $C_{(\alpha_0, \alpha_1, \cdots)}$ as $\hat{u}_j(z)$ and $\hat{v}_j(z)$ $(j = 0,1,\cdots)$ satisfying $\hat{u}(z) C_{(\alpha_0, \alpha_1, \cdots)} = z\hat{u}(z)$ and $C_{(\alpha_0, \alpha_1, \cdots)} \hat{v}(z) = z\hat{v}(z)$, respectfully, where, $\hat{u}(z) = [\hat{u}_0(z) \quad \hat{u}_1(z) \quad \cdots]$, and $\hat{v}(z) =^T [\hat{v}_0(z) \quad \hat{v}_1(z) \quad \cdots]$. Note that if the parameter $(\alpha_0, \alpha_1, \alpha_2, \cdots)$ is changed to $(\alpha_0 e^{iw}, \alpha_1 e^{2iw}, \alpha_2 e^{3iw}, \cdots)$, then the corresponding Laurent polynomials and spectral measure can be rewritten as follows:

$\hat{u}_{2k-1}(z) \rightarrow e^{ikw} \hat{u}_{2k-1}(e^{-ikw} z)$, $\hat{u}_{2k}(z) \rightarrow e^{-ikw} \hat{u}_{2k}(e^{-ikw} z)$, $\hat{v}_{2k-1}(z) \rightarrow e^{-ikw} \hat{u}_{2k-1}(e^{-ikw} z)$

$\hat{v}_{2k}(z) \rightarrow e^{ikw} \hat{u}_{2k}(e^{-ikw} z)$, $\mu(z) \rightarrow \mu(e^{-ikw} z)$.

Let $\alpha \in C$ with $|\alpha| < 1$ and

$$C(\alpha) = \begin{bmatrix} 1-|\alpha|^2 & -\alpha\sqrt{1-|\alpha|^2} & -\alpha\sqrt{1-|\alpha|^2} & \alpha^2 \\ \bar{\alpha}\sqrt{1-|\alpha|^2} & 1-|\alpha|^2 & -|\alpha|^2 & -\alpha\sqrt{1-|\alpha|^2} \\ \bar{\alpha}\sqrt{1-|\alpha|^2} & -|\alpha|^2 & 1-|\alpha|^2 & -\alpha\sqrt{1-|\alpha|^2} \\ (\bar{\alpha})^2 & \bar{\alpha}\sqrt{1-|\alpha|^2} & \bar{\alpha}\sqrt{1-|\alpha|^2} & 1-|\alpha|^2 \end{bmatrix}$$

From the definition of the weight of a passage, the CMV matrix associated with the Verblunsky parameter $(\alpha_0, \alpha_1, \cdots)$, and $U$ for $\Delta = 1$, $\gamma_1 = \gamma_2 = 0$, it is important to note that

$$\Xi^{(I.C(\alpha))}_{(x_2,y_2),(x_1,y_1)}(t) = \begin{bmatrix} (C_{(\alpha,0,\alpha,0,\cdots)})_{(2x_1,2y_1),(2x_2,2y_2)} & (C_{(\alpha,0,\alpha,0,\cdots)})_{(2x_1,2y_1),(2x_2+1,2y_2)} & (C_{(\alpha,0,\alpha,0,\cdots)})_{(2x_1+1,2y_1),(2x_2,2y_2)} & (C_{(\alpha,0,\alpha,0,\cdots)})_{(2x_1+1,2y_1),(2x_2+1,2y_2)} \\ (C_{(\alpha,0,\alpha,0,\cdots)})_{(2x_1,2y_1),(2x_2,2y_2+1)} & (C_{(\alpha,0,\alpha,0,\cdots)})_{(2x_1,2y_1),(2x_2+1,2y_2+1)} & (C_{(\alpha,0,\alpha,0,\cdots)})_{(2x_1+1,2y_1),(2x_2,2y_2+1)} & (C_{(\alpha,0,\alpha,0,\cdots)})_{(2x_1+1,2y_1),(2x_2+1,2y_2+1)} \\ (C_{(\alpha,0,\alpha,0,\cdots)})_{(2x_1,2y_1+1),(2x_2,2y_2)} & (C_{(\alpha,0,\alpha,0,\cdots)})_{(2x_1,2y_1+1),(2x_2+1,2y_2)} & (C_{(\alpha,0,\alpha,0,\cdots)})_{(2x_1+1,2y_1+1),(2x_2,2y_2)} & (C_{(\alpha,0,\alpha,0,\cdots)})_{(2x_1+1,2y_1+1),(2x_2+1,2y_2)} \\ (C_{(\alpha,0,\alpha,0,\cdots)})_{(2x_1,2y_1+1),(2x_2,2y_2+1)} & (C_{(\alpha,0,\alpha,0,\cdots)})_{(2x_1,2y_1+1),(2x_2+1,2y_2+1)} & (C_{(\alpha,0,\alpha,0,\cdots)})_{(2x_1+1,2y_1+1),(2x_2,2y_2+1)} & (C_{(\alpha,0,\alpha,0,\cdots)})_{(2x_1+1,2y_1+1),(2x_2+1,2y_2+1)} \end{bmatrix}$$

$$\Xi^{(II.C(\alpha))}_{(x_2,y_2),(x_1,y_1)}(t) = \begin{bmatrix} (C_{(\alpha,0,\alpha,0,\cdots)})_{(2x_2-1,2y_2-1),(2x_1-1,2y_1-1)} & (C_{(\alpha,0,\alpha,0,\cdots)})_{(2x_2-1,2y_2-1),(2x_1,2y_1)} & (C_{(\alpha,0,\alpha,0,\cdots)})_{(2x_2-1,2y_2),(2x_1-1,2y_1-1)} & (C_{(\alpha,0,\alpha,0,\cdots)})_{(2x_2-1,2y_2),(2x_1,2y_1)} \\ (C_{(\alpha,0,\alpha,0,\cdots)})_{(2x_2-1,2y_2-1),(2x_1,2y_1)} & (C_{(\alpha,0,\alpha,0,\cdots)})_{(2x_2-1,2y_2-1),(2x_1,2y_1)} & (C_{(\alpha,0,\alpha,0,\cdots)})_{(2x_2-1,2y_2),(2x_1,2y_1)} & (C_{(\alpha,0,\alpha,0,\cdots)})_{(2x_2-1,2y_2),(2x_1,2y_1)} \\ (C_{(\alpha,0,\alpha,0,\cdots)})_{(2x_2,2y_2-1),(2x_1-1,2y_1-1)} & (C_{(\alpha,0,\alpha,0,\cdots)})_{(2x_2,2y_2-1),(2x_1-1,2y_1)} & (C_{(\alpha,0,\alpha,0,\cdots)})_{(2x_2,2y_2),(2x_1-1,2y_1-1)} & (C_{(\alpha,0,\alpha,0,\cdots)})_{(2x_2,2y_2),(2x_1-1,2y_1)} \\ (C_{(\alpha,0,\alpha,0,\cdots)})_{(2x_2,2y_2-1),(2x_1,2y_1-1)} & (C_{(\alpha,0,\alpha,0,\cdots)})_{(2x_2,2y_2-1),(2x_1,2y_1)} & (C_{(\alpha,0,\alpha,0,\cdots)})_{(2x_2,2y_2),(2x_1,2y_1-1)} & (C_{(\alpha,0,\alpha,0,\cdots)})_{(2x_2,2y_2),(2x_1,2y_1)} \end{bmatrix}$$

where if $i < j$, then we put $(C^t_{(\alpha_0,\alpha_1,\cdots)})_{ij} = 0$. We should remark that in matrices $\Xi^{(I.C(\alpha))}_{(x_2,y_2),(x_1,y_1)}(t)$ and $\Xi^{(II.C(\alpha))}_{(x_2,y_2),(x_1,y_1)}(t)$, it is understood that $(C_{(\alpha_0,\alpha_1,\cdots)})_{ij} \equiv (C^t_{(\alpha_0,\alpha_1,\cdots)})_{ij}$. The weight of a passage with a general quantum coin $U$, $\Xi^{(J.U)}_{(x_2,y_2),(x_1,y_1)}(t)$, is given by

$$\Xi^{(I,U)}_{(x_2,y_2),(x_1,y_1)}(t) = \left(\Delta^{\frac{1}{2}}\right)^t e^{i\left(\sqrt{(x_2-x_1)^2+(y_2-y_1)^2}\right)\phi} \times D^*(\phi) \Xi^{(I,C(a))}_{(x_2,y_2),(x_1,y_1)} D(\phi)$$

$$\Xi^{(II,U)}_{(x_2,y_2),(x_1,y_1)}(t) = \left(\Delta^{\frac{1}{2}}\right)^t e^{i\left(\sqrt{(x_2-x_1)^2+(y_2-y_1)^2}\right)\phi} \times D^*(\psi) \Xi^{(I,C(a))}_{(x_2,y_2),(x_1,y_1)} D(\psi)$$

where $\psi = (\sigma_R + \sigma_u) - (\gamma_1 + \gamma_2)$, $\phi = \dfrac{(\sigma_R + \sigma_u) - (\sigma_D + \sigma_L)}{2}$, $D(\theta) = diag(e^{i\theta}, 1, 1, e^{-i\theta})$,

$$a = \left(\sum_{\substack{m,n\in\{L,R,D,U\}\\m\neq n}}\bar{c}_{mn}\right)\Delta^{\frac{1}{2}}, \ b = \left(\sum_{\substack{m,n\in\{L,R,D,U\}\\m\neq n}}\bar{c}_{mn}\right)\Delta e^{-i(\gamma_1+\gamma_2)}, \text{ with } \Delta = \det U$$

### III. Spectral measure of CMV matrices and localization

In this section we will define localization as follows: there exists $\Psi^{(J)} \in H^{(J)}$ such that

$\limsup_{t\to\infty}\left|\left\langle \Psi^{(J)}, (W^{(J,U)})^t \Psi_0^{(J)}\right\rangle\right| > 0$ ($J \in \{I, II\}$). Let $\mu^{(I)}$ and $\mu^{(II)}$ be the spectral measures for

$C_{(a,0,a,0,...)}$ and $C_{(0,b,0,b,...)}$ respectively. Let $\{\hat{x}_j(z)\}_{j=0}^{\infty}$ and $\{\hat{X}_j(z)\}_{j=0}^{\infty}$ denote the Laurent polynomials of $\mu^{(I)}$ and $\mu^{(II)}$ satisfying $C_{(a,0,a,0,...)}\hat{x}(z) = z\hat{x}(z)$ and $\hat{X}(z)C_{(0,b,0,b,...)} = z\hat{X}(z)$, where $x(z) = ^T[\hat{x}_0(z) \ \hat{x}_1(z) \ \hat{x}_2(z) \ \cdots]$ and $\hat{X}(z) = \begin{bmatrix}\hat{X}_0(z)\\\hat{X}_1(z)\\\hat{X}_2(z)\\\vdots\end{bmatrix}^T$. Therefore,

$$\left(C_{(a,0,a,0,...)}^t\right)_{lm} = \int_{|z|=1} z^t \hat{x}_l(z)\hat{x}_m(z)d\mu^{(I)}(z), \ \left(C_{(0,b,0,b,...)}^t\right)_{lm} = \int_{|z|=1} z^t \hat{X}_l(z)\hat{X}_m(z)d\mu^{(II)}(z)$$

To get the spectral measure $\mu^{(J)}$ we compute the Caratheodory function $F^{(J)}(z)$

($J \in \{L, R, U, D\}$) which are given by

$F^{(I)}(z) = \lim_{j\to\infty}\frac{\tilde{x}_j(z)}{x_j(z)}$, $F^{(II)}(z) = \lim_{j\to\infty}\frac{\tilde{X}_j(z)}{\tilde{X}_j(z)}$ ($|z| < 1$), where $\tilde{x}_j(z)$ and $\tilde{X}_j(z)$ are the Laurent

polynomials whose Verblunsky parameters are $-\alpha_j$ when the original ones are $\alpha_j$. Put

$B^{(J)} = \{\theta \in [-\pi,\pi)^2 : \lim_{r\uparrow 1} F^{(J)}(re^{i\theta}) = \infty\}$ for $J \in \{L, R, D, U\}$, then the spectral measure is obtained by $d\mu^{(J)}(e^{i\theta}) = w^{(J)}(\theta)\frac{d\theta}{2\pi} + \sum_{\theta_0 \in B^{(J)}} m_0^J(\theta_0)\delta(\theta_0 - \theta)d\theta$, where

$w^{(J)}(\theta) = \lim_{r\uparrow 1}\text{Re}(F^{(J)}(re^{i\theta}))$, $m_0^{(J)}(\theta_0) = \lim_{r\uparrow 1}\frac{1-r}{2}(F^{(J)}(re^{i\theta_0}))$. Here $\text{Re}(z)$ is the real

part of the complex number $z$. Let us write $\theta_{R,L,U,D} = \dfrac{(\sigma_R + \sigma_U) - (\sigma_L + \sigma_D)}{2}$, and

$\hat{x}_i(e^{i\theta_0})\overline{\hat{x}}_j(e^{i\theta_9}) = \hat{x}_i\overline{\hat{x}}_j(e^{i\theta_9})$, then we see that

$$\Xi^{(I)}{}_{(k,k),(0,o)}(t) \sim \left(\Delta^{\frac{1}{2}}\right)^t e^{ik\theta_{R.L.U.D}} \times \sum_{\theta_0 \in B^{(I)}} m_0^I(\theta_0) \begin{bmatrix} \hat{x}_{2k}(e^{i\theta_0}) & \hat{x}_{2k}\overline{\hat{x}}_1(e^{i\theta_0})e^{-i\theta_{R,L,U,D}} & \hat{x}_{2k}\overline{\hat{x}}_1(e^{i\theta_0})e^{-i\theta_{R,L,U,D}} & \hat{x}_{2k}\overline{\hat{x}}_1(e^{i\theta_0}) \\ \hat{x}_{2k+1}(e^{i\theta_0}) & \hat{x}_{2k}\overline{\hat{x}}_1(e^{i\theta_0})e^{-i\theta_{R,L,U,D}} & \hat{x}_{2k+1}\overline{\hat{x}}_1(e^{i\theta_0})e^{i\theta_{R,L,U,D}} & \hat{x}_{2k+1}\overline{\hat{x}}_1(e^{i\theta_0}) \\ \hat{x}_{2k}(e^{i\theta_0}) & \hat{x}_{2k}\overline{\hat{x}}_1(e^{i\theta_0})e^{i2\theta_{R,L,U,D}} & \hat{x}_{2k}\overline{\hat{x}}_1(e^{i\theta_0})e^{-i\theta_{R,L,U,D}} & \hat{x}_{2k}\overline{\hat{x}}_1(e^{i\theta_0}) \\ \hat{x}_{2k+1}(e^{i\theta_0})e^{i2\theta_{R,L,U,D}} & \hat{x}_{2k+1}\overline{\hat{x}}_1(e^{i\theta_0})e^{i\theta_{R,L,U,D}} & \hat{x}_{2k+1}\overline{\hat{x}}_1(e^{i\theta_0})e^{i\theta_{R,L,U,D}} & \hat{x}_{2k+1}\overline{\hat{x}}_1(e^{i\theta_0}) \end{bmatrix}$$

Now write $(\sigma_R + \sigma_U) - (\gamma_1 + \gamma_2) = \theta_{R,U,1,2}$, then we see that

$$\Xi^{(II)}{}_{(k,k),(0,o)}(t) \sim \left(\Delta^{\frac{1}{2}}\right)^t e^{ik\theta_{R.L.U.D}} \times \sum_{\theta_0 \in B^{(I)}} m_0^{II}(\theta_0) \begin{bmatrix} e^{-2i\theta_{R,U,1,2}} \hat{X}_{2k-1}(e^{i\theta_0}) \\ e^{-i\theta_{R,U,1,2}} \hat{X}_{2k}(e^{i\theta_0}) \\ e^{-i\theta_{R,U,1,2}} \hat{X}_{2k-1}(e^{i\theta_0}) \\ \hat{X}_{2k}(e^{i\theta_0}) \end{bmatrix}, \text{ where } A(t) \sim B(t) \text{ means}$$

$\lim_{t \to \infty} \left| \dfrac{(A(t))_{l,m}}{(B(t))_{l,m}} \right| \to 1$, for $l,m \in \{1,2,3,4\}$. Here $A(t)$ and $B(t)$ are $4 \times 4$ matrices.

### IV. Limit measures of the quantum walk

In this section we give explicit expressions for the limit measures for both types of quantum walks considered in this paper.

**Theorem 1:** Let $(X_t^{(I)}, Y_t^{(I)})$ be the Type I quantum walk whose Verblunsky parameter is $(a, 0, a, 0, \ldots)$ at time $t$ with the initial coin state $[\alpha \ \beta \ \mu \ \zeta]^T$ starting from the origin. The quantum coin is given by $U = \begin{bmatrix} c_{RR} & c_{LR} & c_{UR} & c_{DR} \\ c_{RL} & c_{LL} & c_{UL} & c_{DL} \\ c_{RU} & c_{LU} & c_{UU} & c_{DU} \\ c_{RD} & c_{LD} & c_{UD} & c_{DD} \end{bmatrix}$, with $a = \left( \displaystyle\sum_{\substack{m,n \in \{L,R,D,U\} \\ m \ne n}} \overline{c}_{mn} \right)^{\frac{1}{2}} \Delta^{\frac{1}{2}}$. Then we have

$\lim_{t \to \infty} P((X_t^{(I)}, Y_t^{(I)}) = (x, y)) = \dfrac{\operatorname{Re}(a)^2}{1 - \operatorname{Im}(a)^2} \left| \alpha e^{i\theta} + v_I(a)(\mu + \beta) + \zeta v_I(a) e^{-i\theta} \right|^2 (1 + 2v_I^2(a))(v_I^{2x}(a) + v_I^{2y}(a))$

where $\theta = \dfrac{(\sigma_R + \sigma_U) - (\sigma_L + \sigma_D)}{2}$, and $v_I(a) = \dfrac{\mathrm{sgn}(\mathrm{Re}(a))}{\rho}\left(\sqrt{1 - \mathrm{Im}(a)^2} - |\mathrm{Re}(a)|\right)$. Here

$\sigma_i = \arg(c_{ii})$ for $i \in \{L, R, D, U\}$, $\Delta = \det U$ and $\rho = \sqrt{1 - |a|^2}$.

**Proof:** Explicit expressions for the spectral measures and corresponding Laurent polynomials can be shown in a similar way to Appendix A of Konno and Segawa [25], then using them in the asymptotic result for $\Xi^{(I)}{}_{(k,k),(0,0)}(t)$ gives the desired conclusion.

From this theorem we see that the necessary and sufficient condition for Localization is given by the following.

**Corollary 2:** Localization of Type I quantum walk from the origin with initial state $[\alpha \ \beta \ \mu \ \zeta]^T$ occurs if and only if $\mathrm{Re}(a) \neq 0$, $\alpha e^{i\theta} + v_I(a)(\mu + \beta) + \zeta v_I(a) e^{-i\theta} \neq 0$

Concerning the result for Type II quantum walks, we have the following.

**Theorem 3:** Let $(X_t^{(II)}, Y_t^{(II)})$ be the Type II quantum walk whose Verblunsky parameter is $(0, b, 0, b, \ldots)$ at time $t$ with the initial coin state $[e^{i\delta_1} \ e^{i\delta_2}]^T$ starting from the origin. The quantum coin is given by $U = \begin{bmatrix} c_{RR} & c_{LR} & c_{UR} & c_{DR} \\ c_{RL} & c_{LL} & c_{UL} & c_{DL} \\ c_{RU} & c_{LU} & c_{UU} & c_{DU} \\ c_{RD} & c_{LD} & c_{UD} & c_{DD} \end{bmatrix}$, with $b = \left(\displaystyle\sum_{\substack{m,n \in \{L,R,D,U\} \\ m \neq n}} \overline{c}_{mn}\right) \Delta e^{-i(\gamma_1 + \gamma_2)}$. The weight of right or upward moving from the origin is given by $e^{i(\gamma_1 + \gamma_2)}$ ($\gamma_1, \gamma_2 \in \Re$). Then we have

$$\lim_{t \to \infty} P\!\left((X_t^{(I)}, Y_t^{(I)}) = (x, y)\right) = \dfrac{1 + (-1)^{x+y+t}}{2} \times \begin{cases} |M(b)|^2 ; (x, y) = (0,0) \\ |M(b)|^2\left(1 + \dfrac{1}{2v_{II}^2(b)}\right)\!\left(v_{II}^{2x}(b) + v_{II}^{2y}(b)\right); (x, y) \neq (0,0) \text{ and } x, y > 0 \end{cases}$$

where $v_{II}(b) = \dfrac{\rho}{|1 + b|}$, and $M(b) = \dfrac{\left(1 + \mathrm{sgn}(|b|^2 + \mathrm{Re}(b))\right)|b|^2 + \mathrm{Re}(b)}{|(1 + b)^2|}$. Here $\Delta = \det U$ and

$\rho = \sqrt{1 - |b|^2}$.

**Proof:** Explicit expressions for the spectral measures and corresponding Laurent polynomials can be shown in a similar way to Appendix B of Konno and Segawa [25], then using them in the asymptotic result for $\Xi^{(II)}_{(k,k),(0,0)}(t)$ gives the desired conclusion.

From this theorem we see that the necessary and sufficient condition for Localization is given by the following.

**Corollary 4:** Define $D = \{(x, y) \in \mathfrak{R}^2 : x^2 + y^2 < 1\}$. Let the Verblunsky parameter be given by $(0, b, 0, b, \ldots)$ where $b = x + iy$ with $(x, y) \in D$. Then the necessary and sufficient condition for localization of Type II quantum walk is given by

$$(x, y) \in \{(x, y) \in D : \left(x + \frac{1}{2}\right)^2 + \left(y + \frac{1}{2}\right)^2 > \left(\frac{1}{\sqrt{2}}\right)^2\}$$

**V. Summary and Open Problem**

We investigated localization and the limit distribution for quantum walks defined in this paper using the CGMV method. The spectral measure of the CMV matrix corresponding to the quantum walk is given. From the point mass of the measure, localization is shown. We obtained the necessary and sufficient condition for localization with respect to the quantum coin and initial state. In addition, the limit distribution is presented.

The open problem is as follows: What is the relationship between the quantum walk in the quarter plane and the quantum walk on homogeneous trees, if any? Are the properties of the quantum walk in the quarter plane, the same on homogeneous trees?

Concerning the open problem, Konno and Segawa [25] have shown the equivalence between the quantum walk on the half line and the quantum walk on homogeneous trees, and use it to give another proof of localization of quantum walks on homogeneous trees, Corollary 3 of their paper.